\documentclass{ifacconf}

\usepackage{graphicx}      
\usepackage{natbib}        

\usepackage{amsmath}
\usepackage{amssymb}
\usepackage{tabularx}
\usepackage{booktabs}
\usepackage{caption}

\usepackage{color}

\begin{document}
\begin{frontmatter}

\title{Characterizing Driver Interactions with Autonomous Vehicles via Response Maps\thanksref{footnoteinfo}} 
\thanks[footnoteinfo]{This material is based upon work supported by the National Science Foundation under NSF Grant Numbers 2227338, 2345083, and IIS-2107111. Any opinions, findings, and conclusions or recommendations expressed in this material are those of the authors and do not necessarily reflect the views of the National Science Foundation.}

\author[First]{Dave Broaddus} 
\author[First]{Rachel DiPirro} 
\author[Second]{Chishang (Mario) Yang}
\author[First]{Dan Calderone}
\author[Second]{Wendy Ju}
\author[First]{Meeko Oishi}

\address[First]{University of New Mexico, 
   Albuquerque, NM 87131 USA \\(e-mail: \{dbroaddus, rdipirro, dcalderone, oishi\}@ unm.edu)}
\address[Second]{Cornell Tech, 
   New York, NY 10044 USA\\ (e-mail: \{cy546, wendyju\}@cornell.edu)}

\begin{abstract}                
Understanding human responses to autonomous vehicle (AV) behaviors is essential for socially aware interaction, which is crucial for socially compatible navigation in shared traffic environments.
We characterize human driving responses in interactions with AVs as feedback laws over the coupled state space of the human driven vehicle and the AV.
We model the human driver's actions using a response map, a concept based in game theory, and employ a linear representation to capture driver behaviors as a function of AV behaviors, based on empirical data from a driving simulator study.
Our results show that 1) human driver acceleration behavior can be captured using response maps, and 2) human driver responses differ significantly with respect to AV behaviors of yielding, non-yielding, and responsive to the human driver.


\end{abstract}


\end{frontmatter}

\section{Introduction}
Autonomous vehicles (AVs) are increasingly common on US roadways, with companies such as Waymo touting 400,000 weekly trips in six major metropolitan areas (\cite{Waymo}).
However, they can be critically hobbled by a lack of social awareness.
Among many examples, the halting problem has been well-documented by \cite{Brown_Broth_Vinkhuyzen_2023}: AVs sometimes fail to yield when they should, and other times fail to proceed when yielded to.
Addressing and preventing these types of incidents requires new methods and tools that can reason about human response to autonomous vehicle actions, which are in turn dependent upon human actions, and naturalistic human behaviors, which often confound modeling attempts from human factors or other ``first principles'' approaches from cognitive science.
Motivated by these challenges, this paper focuses on data-driven methods to characterize feedback laws that capture naturalistic human response to autonomous vehicles in simulated driving scenarios. 

Building social awareness into autonomous vehicles requires characterization of driver actions in interactive scenarios: at its simplest, in interactions that involve one other vehicle.  Although models that characterize human driving in isolation have been established for decades, in interactive scenarios, actions of one driver inevitably influence the actions of the other driver, and cannot be decoupled (\cite{mcruer_motor_control, Huang_Pitts_2021, DONMEZ2023100932}).
Game theoretic approaches such as those in \cite{7976363}, \cite{9993337}, \cite{11227143}, and \cite{8430842} capitalize upon this coupled relationship, but typically rely upon assumptions of optimality that are known to be idealistic.  Further, human factors models of naturalistic decision making often defy representation in a manner amenable to controls, and are typically focused on high-level decisions as opposed to low-level motor control (\cite{klein1993recognition}).  
In contrast, the data-driven approaches circumvent the need for explicit behavioral models by learning directly from human driving data (\cite{Jain_Liu_Peters_Fridovich-Keil_Topcu_2026, leung2025learning, Avetisyan02012026, Sadigh_Landolfi_Sastry_Seshia_Dragan_2018}).

We propose a data-driven framework to characterize human response in naturalistic driving interactions, that exploits foundational game theoretic concepts of coupling between the dynamics and the control actions of both players, but that does not explicitly require optimality.  Our approach is based in {\em response maps} (\cite{basar_dncgt}) which
characterize the choices an agent makes during interactions.  
Typically derived from an underlying cost structure, we interpret response maps not in terms of the ``best'' or optimal feedback, but rather as any feedback law which satisfies the structural requirements of dependency upon the states of both players.
Specifically, response maps capture how human actions respond to the behavior of others during interactions.

The main contribution of this paper is {\em the creation and interpretation of response maps that capture how humans react to different types of autonomous vehicles in naturalistic driving interactions}.
We choose to model the human feedback law as a response map largely because it is amenable to being learned from data, does not rely on assumptions of optimality, and can capture human behaviors beyond those typically associated with game theoretic equilibria.  However, this representation is also generalizable to stochastic representations (for variation in human responses) and to set‑valued representations (for indifference among multiple actions).
Further, response maps also allow for analysis of desirable properties such as safety and stability, and even more nebulous properties, such as familiarity and comfort in driving styles.


\section{Coupled Dynamics of Human and AV Driving Interactions}
\label{sec:prob_form}

\begin{figure}
    \centering
    \includegraphics[width=\linewidth]{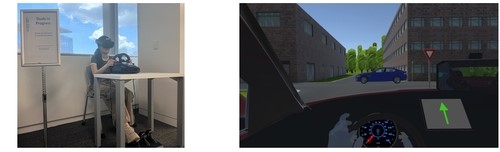}
    \caption{Participant with VR headset (left) and view from inside the driving simulator environment (right)}
    \label{fig:vr-sim}
\end{figure}

\begin{figure}
    \centering
    \includegraphics[width=\linewidth]{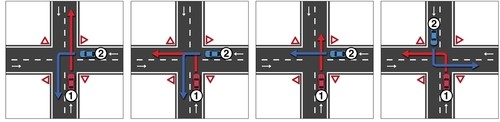}
    \caption{We consider four scenarios, denoted from left to right as S1, S2, S3, and S4. Scenarios S1, S2, and S3 are employed both as shown and with the positions of vehicles 1 and 2 switched, which alters which vehicle has right-of-way. Scenario S4 has ambiguous right-of-way.}
    \label{fig:scenarios}
\end{figure}

\subsection{Experimental Setup}
We focus on an experiment designed to capture interactions between a human driver and three different styles of AVs, described in \cite{Yang_Chang_Dey_Xu_Parush_Ju_2025}.
These data were gathered using the StrangeLand driving simulator (\cite{Goedicke_Zolkov_Friedman_Wise_Parush_Ju_2022}), which enables naturalistic driving scenarios within virtual reality environments (Figure~\ref{fig:vr-sim}).
A driver within this environment controls a simulated vehicle using a physical steering wheel interface, as well as gas and brake pedal controllers.
The driver is able to see and respond to other vehicles within their environment.

Participants navigated four different scenarios at a four-way intersection with yield signs (Figure~\ref{fig:scenarios}): These scenarios involve one or both vehicles making left turns, with right of way for the vehicle to arrive at the intersection, or the vehicle who is on the right, presuming both vehicles arrive simultaneously.
Because right-of-way depends on the direction from which each vehicle approaches the intersection, 
scenarios S1, S2, and S3 were repeated with the vehicles' starting positions swapped.
Scenario S4 involves ambiguous right-of-way.
To distinguish the starting location of each vehicle, scenarios in which the AV has right-of-way are labeled with an ``A,'' and scenarios in which the human-driven vehicle has right-of-way are denoted with a ``B.'' 
In all cases, the velocity of the AV is designed to match that of the participant during their approach, so that both vehicles reach the intersection around the same time.

The AV in this experiment can take on one of the three different types: Contingent, Yield, and NoYield.
The longitudinal behavior of all three AV  models is governed by the Intelligent Driver Model, described in  \cite{Treiber_Kesting_2012}, which is based on a kinematic vehicle model.
The Yield AV always yields before proceeding, continuing through the intersection once it is clear.
The NoYield AV never yields and therefore typically crosses the intersection before the human driver, regardless of right-of-way.
The Contingent AV utilizes a neural network model trained on human-human driving interactions at an intersection to make yielding and acceleration decisions by mimicking observed data.
Specifically, the Contingent model first begins to assess feasibility of crossing the intersection as it approaches by predicting whether it or the human driven vehicle will cross first.
The Contingent AV then continuously assesses the feasibility of crossing via a finite state machine based on its own state and that of the other vehicle until the neural network model determines it is appropriate to cross.


The StrangeLand driving simulator collects data from within the virtual environment, including both vehicle's positions, velocities, accelerations, and headings.
Human experimental data are also collected from the external components, including the angle of the steering wheel, gas and brake pedal information, hand position and orientation, and head position and orientation.
All of the data are collected at approximately 18 frames per second to ensure smooth performance in the simulator, with some variability within trials, and resampled to 20 frames per second for our analysis.

As described in \cite{Yang_Chang_Dey_Xu_Parush_Ju_2025}, 50 New York City-based participants who hold US driver's licenses completed this study. Twenty-four of the participants identified as male, 25 as female, 1 as non-binary. The average age of the participants was 27.5 years, with a standard deviation of 8.3 years.

\begin{figure}
    \centering
    \includegraphics[width=0.99\linewidth]{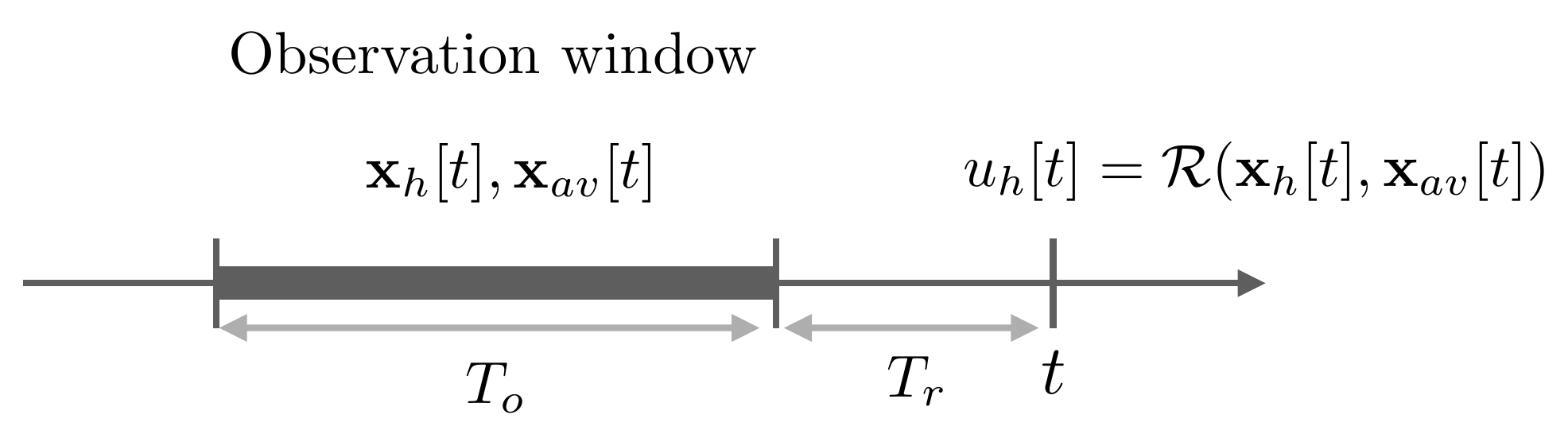}
    \caption{Illustration of the observation window, $T_o$, and the reaction time window, $T_r$.
    }
    \label{fig:timewindows}
\end{figure}

\begin{figure}
    \centering
    \includegraphics[width=\linewidth]{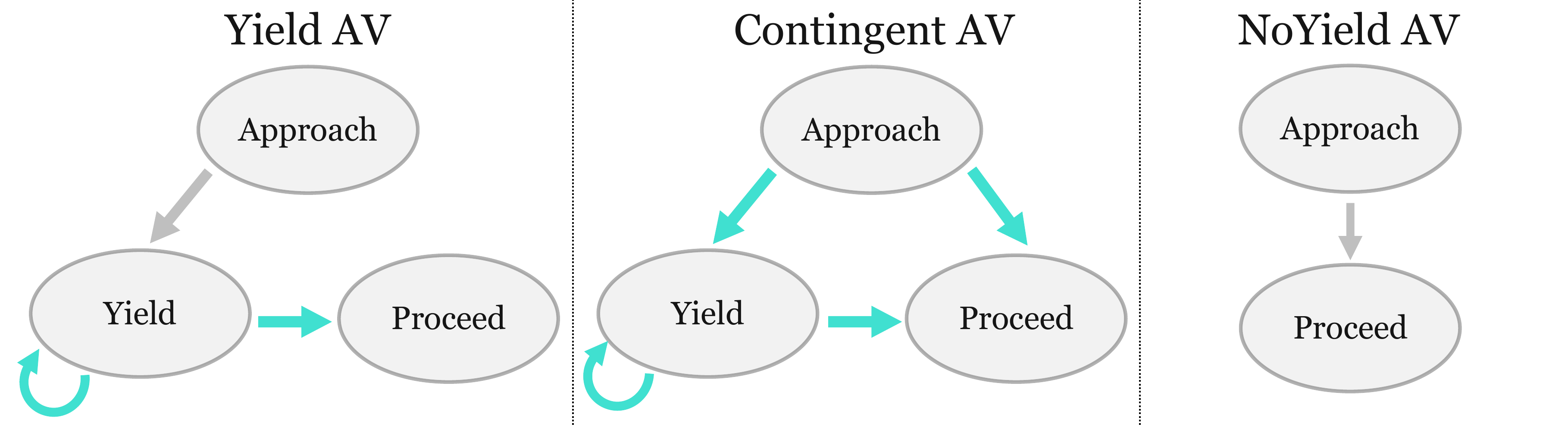}
    \caption{The AV controller is modeled as a hybrid system that depends on AV type. The Contingent type features transitions between modes that are based on the human vehicle state as well as the AV state (blue arrows), whereas the Yield and NoYield types determine whether or not to yield to the human vehicle based only their own state (gray arrows).}
    \label{fig:fsm}
\end{figure}

\subsection{Modeling}

We presume an information processing framework based in engineering psychology (\cite{wickens2021engineering}), in which the human driver uses prior information, gathered within the observation window $T_o$, to make a control decision that is implemented after a reaction time delay, $T_r$.
We presume that $T_o = 3$ seconds from engineering studies in \cite{Lee_Kim_Kim_Yang_2022} and \cite{Eriksson_Stanton_2017} in driving conditions functionally similar to our own, and that $T_r = 0.5$ seconds, from \cite{Huang_Pitts_2021} (Figure~\ref{fig:timewindows}).

We describe the longitudinal and latitudinal coordinates of the human vehicles by $x_h \in \mathbb{R}^2$ and that of the AV by $x_{av} \in \mathbb{R}^2$.
The human driver's input $u_h \in \mathbb{R}$ and the AV's input $u_{av} \in \mathbb{R}$ represent the longitudinal acceleration of human driven and autonomous vehicle, respectively, normalized in the interval $[-1,1]$.

Hence we use
\begin{subequations}
\begin{align}
   \mathbf{x}_{h}[t] & = 
   \begin{bmatrix}
   x_h[t-T_o - T_r]^\top & \dots & x_h[t-T_r]^\top
   \end{bmatrix}^\top 
   \\
   \mathbf{x}_{av}[t] & =
   \begin{bmatrix}
   x_{av}[t-T_o-T_r]^\top & \dots & x_{av}[t-T_r]^\top\end{bmatrix}^\top
\end{align}
\end{subequations}
with $\mathbf{x}_h[t] \in \mathbb{R}^{2T_o}$ and $\mathbf{x}_{av}[t] \in \mathbb{R}^{2T_o}$, to denote the state of the human driven and autonomous vehicles, respectively.

Then, the coupled dynamics of the human driven and autonomous vehicles are described by the dynamics
\begin{subequations}
\begin{align}
\mathbf{x}_h[t+1] & = F_h(\mathbf{x}_h[t],u_h[t]) \label{eq:h-dyanmics} \\
\mathbf{x}_{av}[t+1] & = F_{av}(\mathbf{x}_{av}[t], u_{av}[t]) \label{eq:av-dyanmics}
\end{align}
\end{subequations}
We presume that the functions $F_h$ and $F_{av}$ are not available analytically.

The controller for the AV is dependent upon AV type, and can be modeled for each AV type as a hybrid system with state-triggered transitions, as shown in Figure~\ref{fig:fsm}. The AV type determines the transition conditions: Yield and NoYield AVs transition to ``Yield'' and to ``Proceed'' modes, respectively, based on their position only, whereas the Contingent AV may transition to either of these two modes, as dictated by the output of a neural net trained on human-human driver interactions, which mimics human driver yielding behaviors. The continuous control values in each mode are described in detail in \cite{Yang_Chang_Dey_Xu_Parush_Ju_2025} via the Intelligent Driver Model.

We presume that the controller for the human driven vehicle can be written as the response map
\begin{equation}
\label{eq:human-controller}
    {u}_h[t] = \mathcal{R}(\mathbf{x}_h[t], \mathbf{x}_{av}[t]) \in \mathbb{R}.
\end{equation}

For each trial $i$, we observe the human driven vehicle and AV states and inputs starting at an initial time $t^i_0$ and continuing for $T^i$ time steps, captured via data matrices
\begin{subequations}
    \begin{align}
    S_x^i 
    &= \begin{bmatrix}
        \mathbf{x}_h^i[t_0^i] & \cdots & \mathbf{x}_h^i[t_0^i +T^i] \\
        \mathbf{x}_{av}^i[t_0^i] & \cdots &  \mathbf{x}_{av}^i[t_0^i +T^i]
    \end{bmatrix}  \in \mathbb{R}^{4T_o \times T^i}\\
        S_u^i &= \begin{bmatrix}
            u_h^i[t_0^i] & \cdots & u_h^i[t_0^i + T^i] 
        \end{bmatrix} \in \mathbb{R}^{1 \times T^i}.
    \end{align}
\end{subequations}
The problem we seek to address in this paper is the inference of the response map \eqref{eq:human-controller} from data $S_x^i$ and $S_u^i$ for each of the three AV types.

\begin{table}[t]
    \centering
    \caption{Number of trials per group, \\ Human driven vehicle first scenarios}
    \setlength{\tabcolsep}{5pt} 
    \small
    \begin{tabular}{lccccccc}
    \toprule
       \textbf{Scenario} & \textbf{S1A} & \textbf{S1B} & \textbf{S2A} & \textbf{S2B}& \textbf{S3A} & \textbf{S3B} & \textbf{S4}   \\
    \midrule          
    \textbf{NoYield} & 0 & 1 & 3 & 0 & 0 & 3& 3  \\
    \textbf{Contingent} & 17 & 17 & 10 & 16& 7 & 19 & 30  \\
    \textbf{Yield} & 18 & 22 & 17 & 17& 19 & 22 & 36  \\
    \bottomrule
    \end{tabular}
    \label{tab:countsH}
\end{table}

\begin{table}[t]
    \centering
    \caption{Number of trials per group, \\AV first scenarios}
    \setlength{\tabcolsep}{5pt} 
    \small
    \begin{tabular}{lccccccc}
    \toprule
       \textbf{Scenario} & \textbf{S1A} & \textbf{S1B} & \textbf{S2A} & \textbf{S2B} & \textbf{S3A} & \textbf{S3B} & \textbf{S4}  \\
    \midrule          
    \textbf{NoYield} & 25 & 23 & 18 & 25 & 25 & 22 & 47  \\
    \textbf{Contingent} & 8 & 8 & 15 & 9 & 18 & 6 & 20  \\
    \textbf{Yield} & 7 & 3 & 8 & 8 & 6 & 3 & 13  \\
    \bottomrule
    \end{tabular}
    \label{tab:countsAV}
\end{table}

\section{Methods}
\label{sec: Methods}

We first hypothesize that the mapping $\mathcal{R}(\cdot, \cdot)$ in \eqref{eq:human-controller} can be locally approximated by a linear function, motivated by prior work demonstrating that linear models effectively capture human decision-making in driving tasks (\cite{Mcruer}), as well as studies that have identified 
differences in driver behaviors
through similar linearization approaches (\cite{10919603, 11423689}).
We interpret this linearized response map as describing local variations around a baseline policy of expected driver behavior, to capture subtle differences in acceleration and braking due to vehicle interactions.

Under this linear hypothesis, we infer \eqref{eq:human-controller} from data via regression, by assuming
\begin{equation}
    \mathcal{R}(\mathbf{x}_h[t], \mathbf{x}_{av}[t]) = w_h^\top \mathbf{x}_{h}[t] + w_{av}^\top \mathbf{x}_{av}[t] + b
\end{equation}
with constant, unknown vectors $w_h \in \mathbb{R}^{2T_o}$, $w_{av} \in \mathbb{R}^{2T_o}$, and $b \in \mathbb{R}$.
We calculate these vectors by minimizing the mean squared error,
\begin{equation}
        \min_{w_h, w_{av}, b} \sum_{i=0}^N\sum_{t=t_0^i}^{T^i}||u_h^i[t] - \mathcal{R}(\mathbf{x}_h^i[t],\mathbf{x}_{av}^i[t])||_2^2
    \label{eq:mini}
\end{equation}
which over $N$ trials has the closed-form solution
\begin{equation}
    W = (X E^{-1} X^\top)^{-1} X E^{-1} U_h
\end{equation} for $W=[w_h^\top, w_{av}^\top, b]^\top \in \mathbb{R}^{4T_o+1}$, state data $X = \begin{bmatrix}
    S_x^1 \cdots S_x^N \\
    (\mathbf{1}^1)^\top \cdots (\mathbf{1}^N)^\top 
\end{bmatrix} \in \mathbb{R}^{(4T_o+1) \times s}$ 
with $s= \sum_{i=0}^N T^i$,
input data 
$U_h = \begin{bmatrix}
    S_u^1 \cdots S_u^N
\end{bmatrix}^\top \in \mathbb{R}^{s}$,
and error matrix $E \in \mathbb{R}^{s \times s}$.
We presume $\mathbf{1}^i \in \mathbb{R}^{T^i}$ is a vector of ones.

The linear response map offers practical advantages: it can be learned with fewer samples as compared to a nonlinear model and is less prone to overfitting than a nonlinear model, particularly in regions of the state space where data is limited, and the parameters of the linear model admit direct physical interpretation.
The linear structure is readily amenable to analysis and controller synthesis, as it enables the use of established tools from both linear systems and game theory.

\section{Results}
\label{sec: results}

\subsection{Data Pre-processing}

The data capture 7 scenarios and 3 AV policies, resulting in 42 groups: 21 in which the human driven vehicle entered the intersection first, and 21 in which the AV entered first.
Of these 42 groups, we consider the 31 groups that had sufficient data for analysis.
All 
No Yield groups were excluded for having fewer than 8 trials (Table \ref{tab:countsH}), and 
all scenario S3B groups were excluded because Contingent and Yield had fewer than 8 trials, and therefore could not be compared with the NoYield group (Table \ref{tab:countsAV}). Scenario S1B was also excluded due to only having 3 trials (Table \ref{tab:countsAV}).
Additionally, eight trials were excluded because they did not have enough data to capture the observation and reaction time windows, or because human drivers performed non-prescribed behaviors, such as turning when instructed to go straight or simply not crossing the intersection.

Observed trajectories were initialized when both vehicles were less than 30 meters from the intersection, and ended once the human-driven vehicle and the AV had finished interacting.
For trials in which the human driven vehicle passed through the intersection first, the interaction ended 1 second (20 frames) after the human driven vehicle passed the center of the intersection.
For trials in which the AV passed through the intersection first, the interaction ended 2 seconds (40 frames) after the AV passed the center of the intersection.
The difference in stopping time reflects our expectation that the human driver's response depends on whether they pass in front of the AV or wait until the AV has cleared the intersection.
Since we seek to learn how human drivers react to different AVs, we end the interaction earlier when the human driven vehicles goes first.
That is, once the human driven vehicle has passed the AV, we assume they no longer interact with it.

\begin{figure}[t]
    \centering
\includegraphics[width=\linewidth]{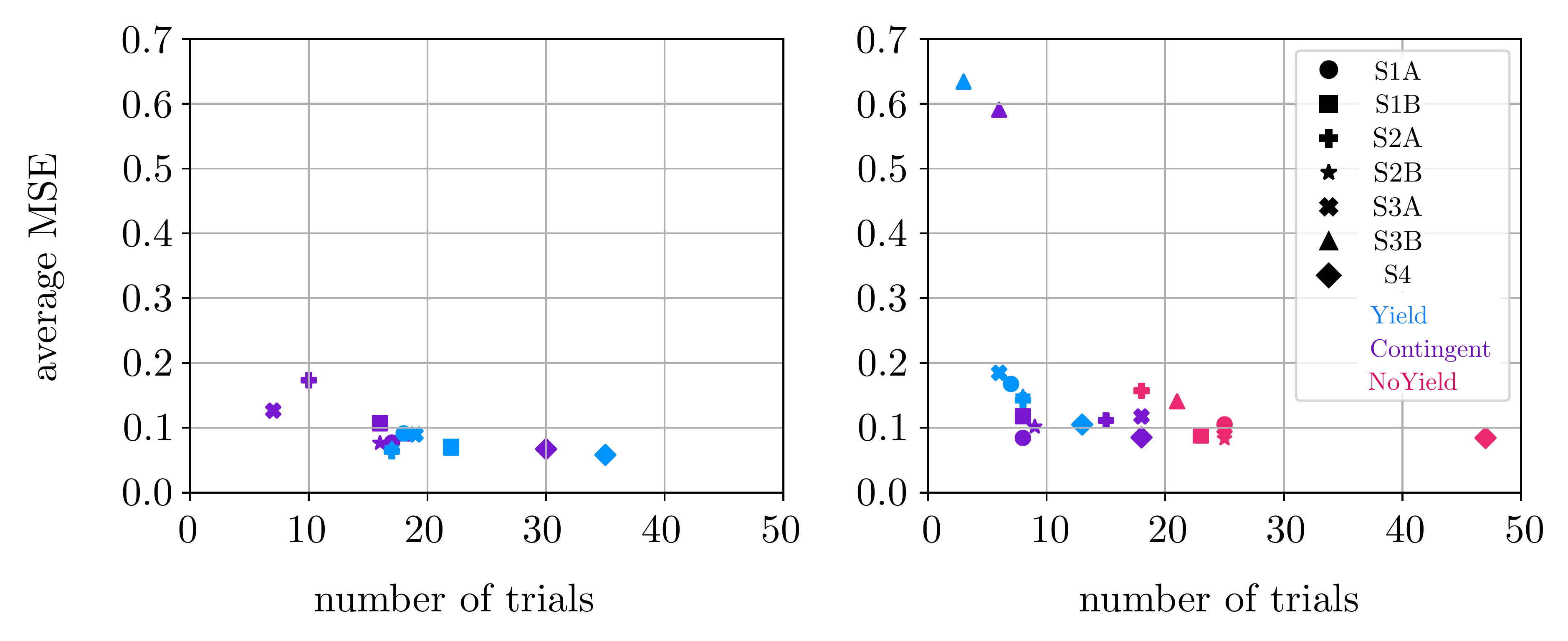}
    \caption{Mean squared error per number of trials for each model. Left: Models in which the human driver crosses the intersection first. Right: Models in which the AV crosses the intersection first. Small magnitudes indicate the model fits the data well.}    
    \label{fig:meanMSEs}
\end{figure}

\subsection{Model Validation}

We assessed the fit of the learned response maps as calculated via \eqref{eq:mini} by performing leave-one-out validation (\cite{hastie01statisticallearning}) 
on the mean squared error between the predicted and observed human driven vehicle input
for each trial in the group.
As shown in Figure 
\ref{fig:meanMSEs}, all 31 models had mean square errors less than 1, well below the the maximum possible value of 4.
Two groups, in which the Yield and Contingent AVs arrived first in scenario S3B (blue and purple triangles, respectively), had relatively  large mean squared error values of 0.63 and 0.59, respectively; this reflects the small number of data points within each of these groups.
Because the generalization error as characterized by the mean squared error is low for the vast majority of our models, we maintain that \eqref{eq:mini} sufficiently fits the data to warrant our next analysis.


\begin{figure}[t]
    \centering
    \includegraphics[width=0.75\linewidth]{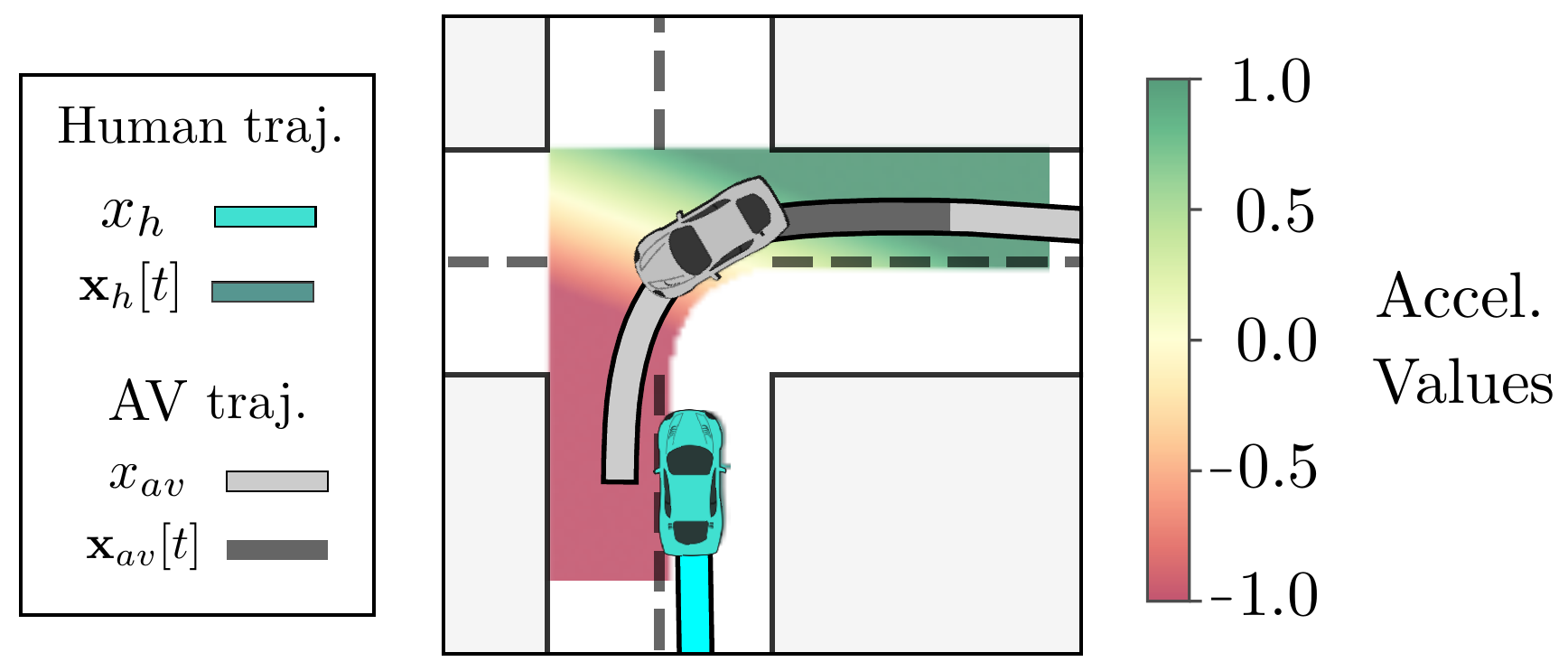}
    \caption{(Scenario S1A, Yield AV) Learned response maps capture expected acceleration of the human driven vehicle (blue) when it is farther from the AV (gray), and braking when it is closer to the AV. 
    }
    \label{fig:gradient_makesense}
\end{figure}

\begin{figure*}[t]
\centering
\includegraphics[width=\linewidth]{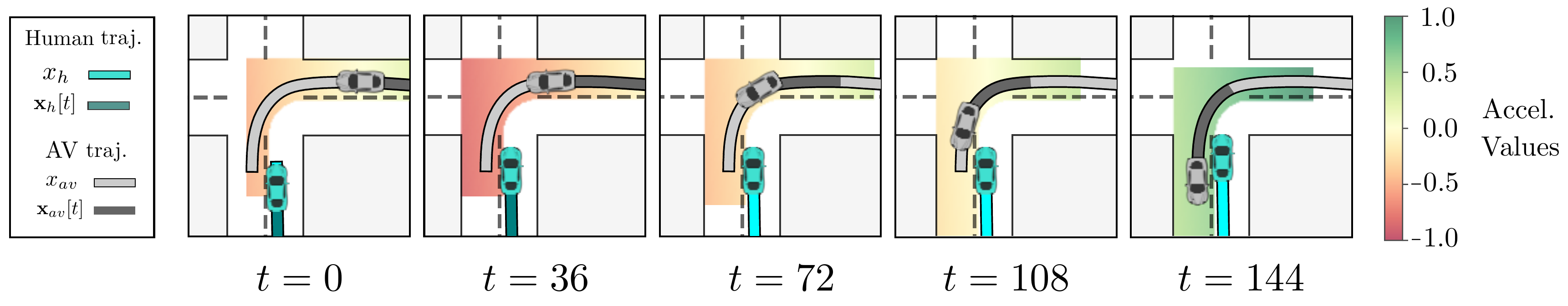}

\caption{For scenario S1A with a Contingent AV, we see that the learned response map predicts expected driver behavior as the Contingent AV (gray) crosses the intersection.  The human driven vehicle brakes with increasing magnitude (red) as the AV nears the intersection, then eases off the brake and begins to accelerate (green) as the human driven vehicle's anticipated path clears.}
\label{fig:expected_behavior}
\end{figure*}




\begin{figure}[t]
\centering
\includegraphics[width=\linewidth]{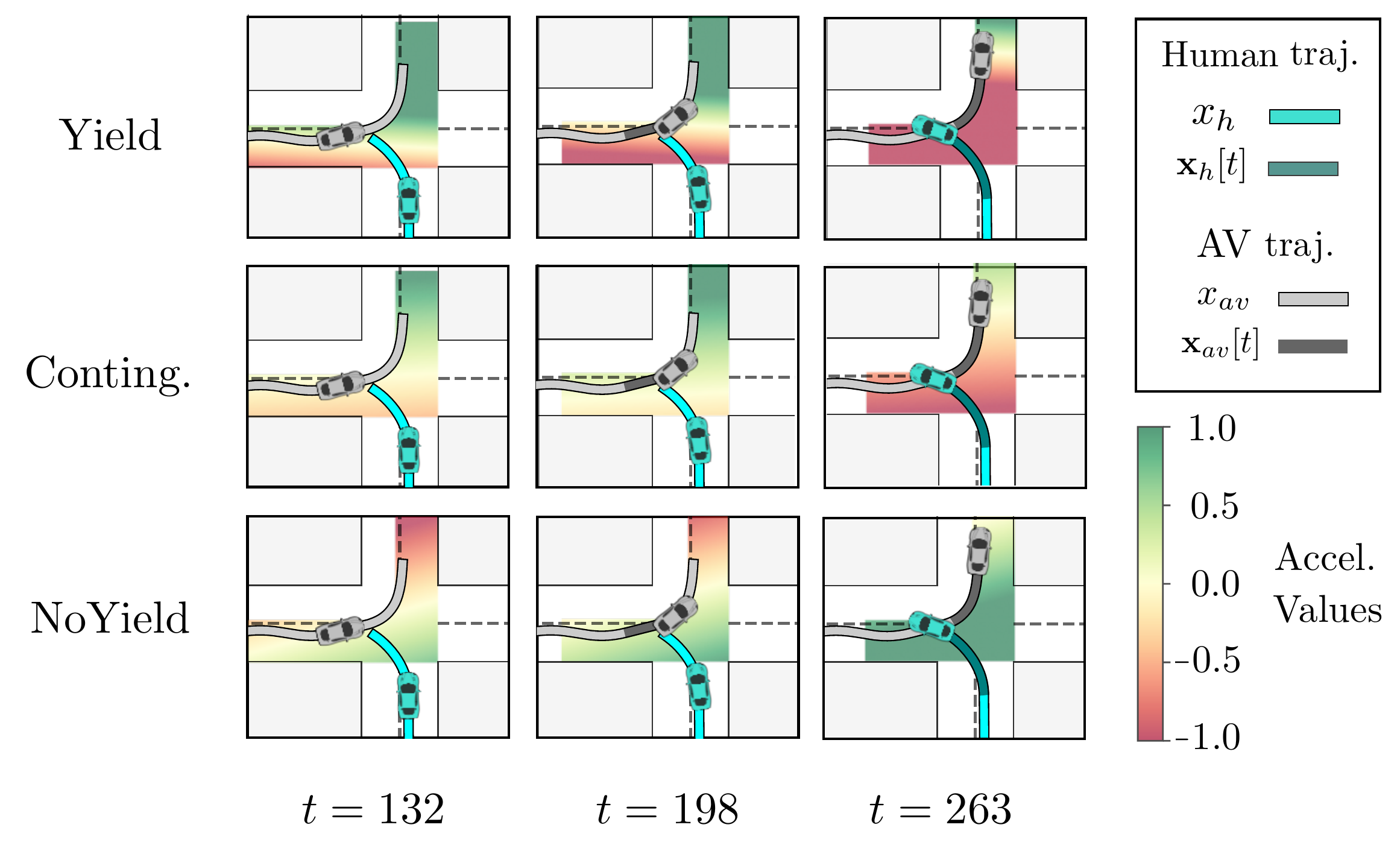}
\caption{(Scenario S2B) With Yield and Contingent AVs, the learned response maps show that the human driven vehicle (blue) brakes when closer to the AV (gray).  With the NoYield AV, the AV passes quickly enough that the human driven vehicle does not need to brake. 
}
\label{fig:scen3B_AVfirst_AVshift}
\end{figure}

\subsection{Human Response Maps}

We analyze and interpret the learned response maps as ``heat maps,'' which show the magnitude of predicted human driven vehicle's acceleration,
obtained by varying the relative position of the AV's trajectory within the state space and computing the human driven vehicle's predicted input from the learned response map.
To do so, we translate the segment of the AV trajectory that generates the response map while holding the trajectory of the human driven vehicle constant. 
As shown in Figure~\ref{fig:gradient_makesense}, 
the color of the heat map indicates the control value that the human driven vehicle would use, if the AV were located at that point at time $t$.  
For this scenario, we see that as expected,
the human driven vehicle accelerates when it is far from the AV, waits when the AV is directly in front of it, and brakes when it is close to the AV.
We focus in this section on interpretation of three specific scenarios in which the AV crosses the intersection first -- one scenario is typical of many of the scenarios analyzed, and the other two scenarios highlight the variability in human acceleration and braking as a function of AV type, when the AV passes in front of the human driver.

\subsubsection{Response maps capture expected driving behaviors.}
A total of 19 of the 31 groups considered, including the one shown in Figure~\ref{fig:expected_behavior} (scenario S1A, Contingent AV), demonstrate the expected acceleration and braking behavior as the human driven vehicle approaches the intersection and waits for the Contingent AV to cross.
At $t=0$, the human driven vehicle begins decelerating.
Then at $t=36$, the AV enters the intersection, so the human driven vehicle brakes, as shown by the red color in the map.
Later at $t=72$ and $t=108$, the human driven vehicle stops braking but does not move, not using either the acceleration nor the brake pedals, as they wait for the AV to cross the intersection.
Finally, at $t=144$, the AV exits the intersection, so the human driven vehicle accelerates, shown by the green color in the map.

\subsubsection{Response maps vary between AV types.}

\begin{figure}
    \centering
    \includegraphics[width=\linewidth]{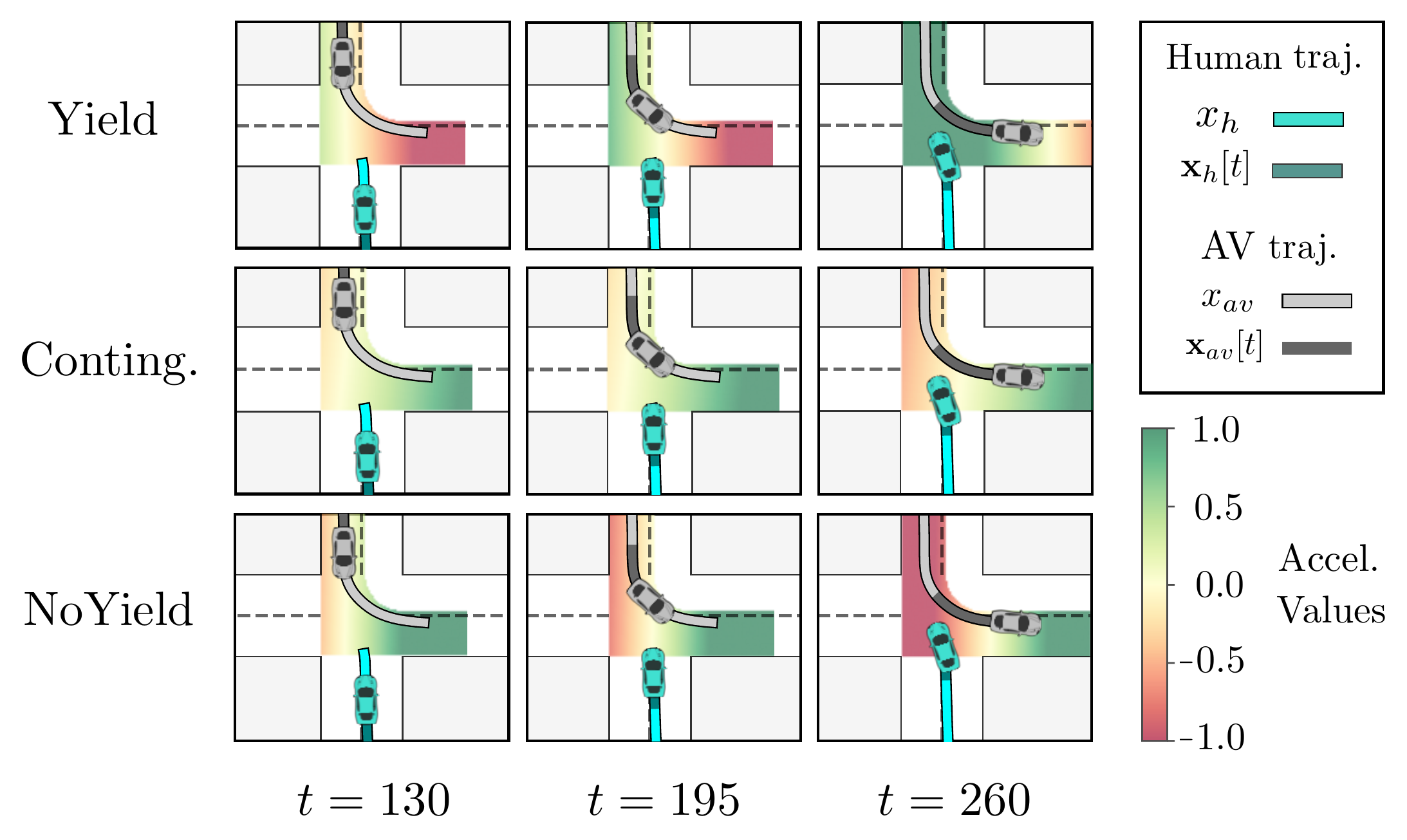}
    \caption{(Scenario S4) 
    With the Yield AV, the learned response map shows that the human driven vehicle (blue) accelerates earlier in the trial than with the Contingent or NoYield AVs.
    }
    \label{fig:scen5_diff_response_maps}
\end{figure}

We next demonstrate differences in the human driven vehicle's responses across AV behavior types, as these distinctions play a central role in shaping interaction outcomes.
In Figure~\ref{fig:scen3B_AVfirst_AVshift} (scenario S2B), we compare the Yield, Contingent, and NoYield response maps for the case in which the AV enters the intersection first.
The Yield and Contingent response maps are similar: when the AV is close to the human driven vehicle, the human driven vehicle brakes.
However, the response to the NoYield differs considerably -- instead of braking, the human vehicle accelerates, anticipating that the NoYield vehicle will move quickly out of the intersection.
In Figure~\ref{fig:scen5_diff_response_maps} (scenario S4), 
the response map for the Yield AV differs starkly from response maps for the Contingent and NoYield AVs.
With the Yield AV, the human driven vehicle accelerates early in the trial, when the AV is to the left of the human driven vehicle, and then waits and brakes, before finally accelerating again.
With the Contingent and NoYield AVs, because the AV does not yield at the intersection, the human driven vehicle yields when the AV is to the left of the human driven vehicle.
Although this scenario has an ambiguous right-of-way,
which technically means that neither vehicle needs to stop for the other, 
the human driver still decelerates when approaching the intersection, perhaps revealing hesitance about the AV's behaviors.

\subsection{Implications for Control}
Our primary finding is that this method can effectively infer control laws underlying human driving behavior that are interpretable functions of AV driving behavior.
In addition to providing insight into human driver response to AVs, this approach 
is also amenable to mathematical analysis (for i.e., safety or stability) via its linear feedback form (although not demonstrated here).  
Our secondary finding is that with learned response maps, 
we demonstrate that human feedback laws are indeed reactive to AV behavior.
Consequently, by modifying the AV's behavior, it is possible to influence how a human driver interacts with an AV.

\section{Conclusions \& Future Work}
\label{sec: conclusions}
We describe a method to infer human feedback laws, in the form of response maps, that are responsive to AV state.  Our approach employs regression to construct response maps from observed data, for a given AV type, in a manner that is dependent on the human driven vehicle state and the AV state (including a recent segment of the trajectory).  One of the main advantages of this form is that because it is a linear control law, it is amenable to control analysis (which is the subject of future work).  Our approach enables coupled quantitative and qualitative interpretation of human-AV interaction in a naturalistic driving simulator.



\section*{DECLARATION OF GENERATIVE AI AND AI-ASSISTED TECHNOLOGIES IN THE WRITING PROCESS}
During the preparation of this work the authors used Copilot (Microsoft) to suggest word choice and grammar for improved overall writing quality. After using this tool, the authors reviewed and edited the content as needed and take full responsibility for the content of the publication.

\bibliography{ifacconf}             
                                                   







\end{document}